\documentclass[9pt,twocolumn,twoside]{opticajnl}
\usepackage{babel}
\usepackage{hyperref}
\journal{opticajournal}
\setboolean{shortarticle}{true} 

\begin{document}

\title{Normal dispersion Kerr cavity solitons:\newline beyond the mean-field limit}
\author[1,2,*]{Thomas G. Seidel}
\author[2]{Julien Javaloyes}
\author[1,2,3]{Svetlana V. Gurevich}

\affil[1]{Institute for Theoretical Physics, University of M\"unster, Wilhelm-Klemm-Str. 9, 48149 M\"unster, Germany}
\affil[2]{Departament de F\'{\i}sica, Universitat de les Illes Balears \& IAC-3, Cra.\,\,de Valldemossa, km 7.5, E-07122 Palma de Mallorca, Spain}
\affil[3]{Center for Nonlinear Science (CeNoS), University of M\"unster, Corrensstraße 2, 48149 M\"unster, Germany}

\affil[*]{Corresponding author: thomas.seidel@uni-muenster.de}
\graphicspath{{./}{./figs/}} 

\begin{abstract}
We predict the existence of a novel type of temporal localized structure in injected Kerr--Gires--Tournois interferometers (KGTI). These bright pulses exist in the normal dispersion regime, yet they do not correspond to the usual scenario of domain wall locking that induces complex shape multistability, weak stability, and a reduced domain of existence. 
The new states are observed beyond the mean-field limit and out of the bistable region. 
Their shape is uniquely defined, with peak intensities beyond that of the upper steady state, and they are stable over a broad range of the injection field, highlighting their potential for optical frequency comb (OFC) generation.
\end{abstract}

\maketitle

Optical frequency combs (OFCs) are coherent light sources consisting of a series of evenly spaced and coherent discrete spectral components. They permit the measurement of light frequencies with exquisite accuracy and possess a plethora of applications in fundamental and applied physics; see, e.g.,~\cite{CY-RMP-03,D-JOSAB-10,PPR-PR-18,FB-CommPhys-19}. Prominent realizations of OFCs leverage the temporal cavity solitons obtained both in injected passive Kerr resonators, such as fiber loops~\cite{LCK-NAP-10} or microrings~\cite{HBJ-NAP-14}, and mode-locked lasers~\cite{WVM-LSA-17}.

In the regime where the intracavity power is relatively low and the overall cavity detuning remains small during propagation, the mean-field approximation can be employed, and the formation of OFCs in Kerr resonators is well-described by the Lugiato--Lefever equation (LLE) or coupled-mode models~\cite{LL-PRL-87,PPR-PR-18,CSY-PRL-10,CY-PRA-10,CM-PRA-13,CRSE-OL-13,HBJ-NAP-14,CPP-PRL-21}. Using the LLE, both bright and dark temporal localized states (TLSs)~\cite{LCK-NAP-10,LLK_OE_15,XXL_NP_15} can be found in the anomalous and normal dispersion regimes, respectively. In the latter case, third-order dispersion also allows for stable dark and bright TLSs to coexist~\cite{PGL-OL-14,PGK_OL_16,PGG_PRA_17}.
However, mean-field models are not always able to account for the complete range of complex dynamics observed and alternative approaches based on, e.g., extended LLE equations~\cite{CB-OL-17} or Ikeda map~\cite{I-OC-79} models are used. The latter allows in particular the accurate theoretical and experimental description of several nonlinear resonances and high-power super cavity solitons associated with the multistable states~\cite{HW-JOSAB-15,AWL-PRX-17}.

Recently, an alternative method for the generation of phase-locked tunable OFCs in optically injected vertical-emission Kerr--Gires--Tournois interferometers (KGTI) has been proposed~\cite{SPV-OL-19,SJG-OL-22}. There, the underlying model derived from first principles relies on delay algebraic equations (DAEs). In particular, in the normal dispersion regime, bright and dark TLSs formed via the locking of domain walls connecting the high- and low-intensity levels of the injected micro-cavity, leading to complex shape multistability.
The obtained TLSs are typically in the picosecond range, thereby resulting in OFCs having THz bandwidth and tunable GHz repetition rate. The latter can be spatially multiplexed in the transverse plane of the microdisk. For small cavity losses and weak injection, the dynamics of KGTI system  can be well-described by the LLE with third-order dispersion~\cite{SGJ-PRL-22}. 

In this Letter, using path continuation methods ~\cite{DDEBT}, we demonstrate the existence of a new type of short high-intensity TLS in the normal dispersion regime, operating the KGTI system far beyond the mean-field limit. These robust TLSs extend beyond the optical bistability region and exhibit a high potential for OFC applications
due to their higher intensity and shorter duration.

The KGTI system is composed of a monomode microdisk cavity of a few micrometers containing a Kerr medium. The micro-cavity is closed by two distributed Bragg mirrors
with reflectivities $r_{1,2}$ and is coupled to a long external cavity of a few centimeters with
round trip time $\tau$ which is closed by a mirror with reflectivity $\eta$. This whole system is subjected to continuous-wave injection with amplitude $Y_0$ and frequency $\omega_0$.
Following the methods developed in~\cite{SPV-OL-19,SJG-OL-22,SGJ-PRL-22,KSG-OL-22}, we can write the evolution of the normalized slowly varying field envelopes in the micro-cavity $E$ and the external cavity $Y$ as follows:
\begin{align}
	\dot{E} &= \left[-1 + i \left(|E|^2-\delta\right)\right] E + h Y \label{eq:DAE1}\,,\\
	Y &= \eta e^{i\varphi} \left[E(t-\tau) - Y(t-\tau)\right] + \sqrt{1-\eta^2}Y_0\,, \label{eq:DAE2}
\end{align}
where $t$ is the time, $\delta=\omega_c-\omega_0$ is the detuning with respect to the
micro-cavity resonance $\omega_c$ and $\varphi=\phi+\omega_0\tau$ is the total phase that allows to lock the
injection to the nearest external cavity mode. It is composed of the phase shift $\phi$ at the external mirror and the accumulated phase when propagating through the external cavity $\omega_0\tau$. The coupling between micro- and external cavity is given by the factor $h=\left(1+|r_2|\right)\left(1-|r_1|\right)/\left(1-|r_1||r_2|\right)$. Note that for a perfectly reflecting bottom mirror $r_2=1$ and, hence, $h=2$, which corresponds to the imbalanced Gires--Tournois interferometer regime~\cite{GT-CRA-64}.
The output $O=E - Y$ is the combination of the intra-cavity photons transmitted and reflected by the micro-cavity and it is re-injected after attenuation $\eta$ and a time delay $\tau$. The coupling between the intra- and external cavity fields is given by a delay algebraic equation (DAE) (Eq.~(\ref{eq:DAE2})) which takes into account all the multiple reflections in a possibly high finesse external cavity. Hence, both group delay dispersion and third-order dispersion are naturally captured by Eqs.~(\ref{eq:DAE1}) and (\ref{eq:DAE2})~\cite{SCM-PRL-19,SPV-OL-19}; see also \cite{VD-PRE-24}.
Note that our modeling approach is similar to the one used for microdisk mode-locked lasers; see e.g., \cite{MB-JQE-05,CSV-OL-18,SCM-PRL-19,SHJ-PRAp-20,HGJ-OL-21}.
%

We start our analysis by calculating the homogeneous steady state (HSS) solutions of Eqs.~(\ref{eq:DAE1}) and (\ref{eq:DAE2}) by plugging Eq.~(\ref{eq:DAE2}) into Eq.~(\ref{eq:DAE1}) and solving for $Y_0$:
\begin{equation}
	Y_{0}^{2}=\frac{1}{h^{2}}\frac{\left|1+\eta e^{i\varphi}\right|^{2}}{1-\eta^{2}}\left|\frac{1-\left(h-1\right)\eta e^{i\varphi}}{1+\eta e^{i\varphi}}-i\left(I_{s}-\delta\right)\right|^{2}I_{s}\,,
	\label{eq:KGTI stst}
\end{equation}
where $I_s=|E|^2$ is the field intensity of the HSS. On the other hand, one can also derive an input--output relation between the injected field $Y_0$ and the output intensity $|O_s|^2$:
\begin{align}
	\left|O_s\right|^{2}&=\frac{\left(1-\eta^{2}\right)Y_{0}^{2}}{\left|1-\eta e^{i\left(\varphi-2\arctan\theta\right)}\right|^{2}}\,,
	\label{eq:input output}
\end{align}
where we define $\theta=\delta-I_s$. 
\begin{figure}[t!]
	\centering
	\includegraphics[width=1\columnwidth]{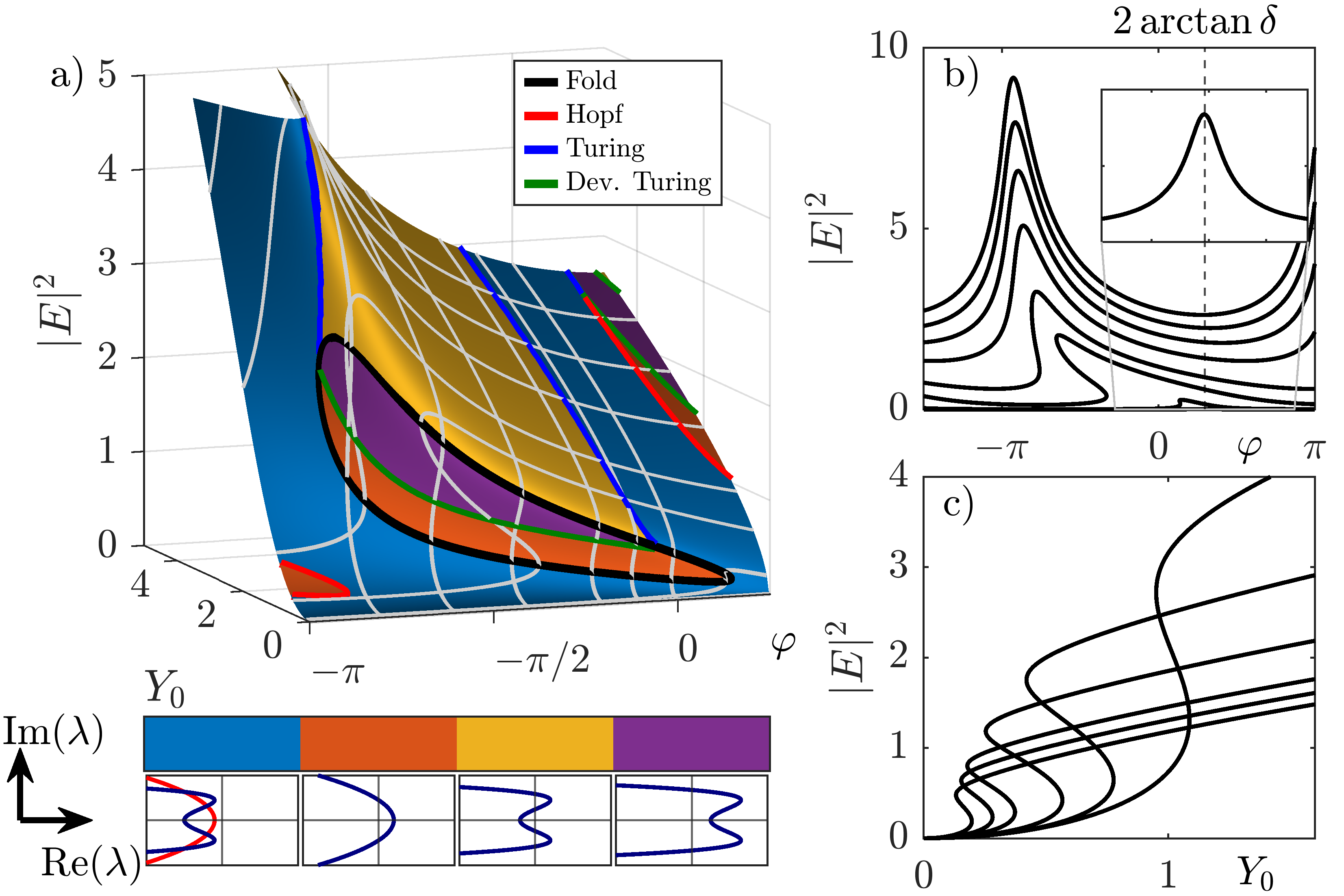}
	\caption{(a) HSSs defined in Eqs.~(\ref{eq:KGTI stst}) and (\ref{eq:input output}) in the $(\varphi,Y_0$)-plane for $(\delta,\,h,\,\eta)=(0.5,\,2,\,0.75)$. The colors encode the stability information as indicated in the legend at the bottom: blue, orange, yellow, and purple correspond to  stable, uniformly unstable, Turing unstable, and developed Turing unstable, respectively. The boundaries between the color patches are bifurcations of the HSSs as indicated by the legend at the top. The white lines are cuts in the one-parameter planes $(\varphi,|E|^2)$ and $(Y_0,|E|^2)$ and are shown in panels (b) and (c), respectively.}
	\label{fig:stability_plane}
\end{figure} 

Now, using Eqs.~(\ref{eq:KGTI stst}) and (\ref{eq:input output}), one can reconstruct the whole set of the HSS solutions as the functions of, e.g., the injection $Y_0$ and the phase $\varphi$, keeping other parameters fixed. The results for the normal dispersion regime, i.e., for $\delta>0$ are summarized in Fig.~\ref{fig:stability_plane}(a) together with the corresponding cuts in $\varphi$- and $Y_0$-directions in panels (b) and (c), respectively.
Here, the cuts in $\varphi$ in Fig.~\ref{fig:stability_plane}(b) corresponds to the resonances of the nonlinear cavity. For low injection values, the intensity $|E|^2$ is small such that $\theta\sim \delta$. In this case Eq.~(\ref{eq:input output}) describes the linear cavity response that has its maximum at $2\arctan\delta$ where the effective phase $\tilde\varphi\equiv\varphi-2\arctan\delta$ vanishes; cf. the inset.

Further, the color map in Fig.~\ref{fig:stability_plane}(a) encodes the linear stability of the HSS solutions. Here, each color corresponds to a qualitatively different form of the pseudo-continuous spectrum~\cite{Yanchuk2005} of eigenvalues $\lambda$ as indicated by the legend: blue, orange, yellow, and purple regions correspond to stable, uniformly unstable, Turing unstable, and Turing unstable with an unstable zero-frequency eigenvalue solutions, respectively. Note that in the long cavity limit, the form of the spectrum is independent of the time delay $\tau$, only the position of the eigenvalues on the spectrum depends on the delay.

Further, colored lines in Fig.~\ref{fig:stability_plane}(a) correspond to the possible bifurcations of the HSS. Here, in particular, the red line located around $\varphi=-\pi$ corresponds to an Andronov--Hopf (AH) bifurcation leading to the regime of the so-called square waves--periodic solutions with a repetition rate that is approximately twice the external cavity round trip~\cite{KSG-OL-22}.
Next, black lines indicate the folds of the HSS. They can be found analytically by calculating the derivative of Eq.~(\ref{eq:KGTI stst}) with respect to $I_s$.

\begin{figure}[t!]
	\centering
	\includegraphics[width=1\columnwidth]{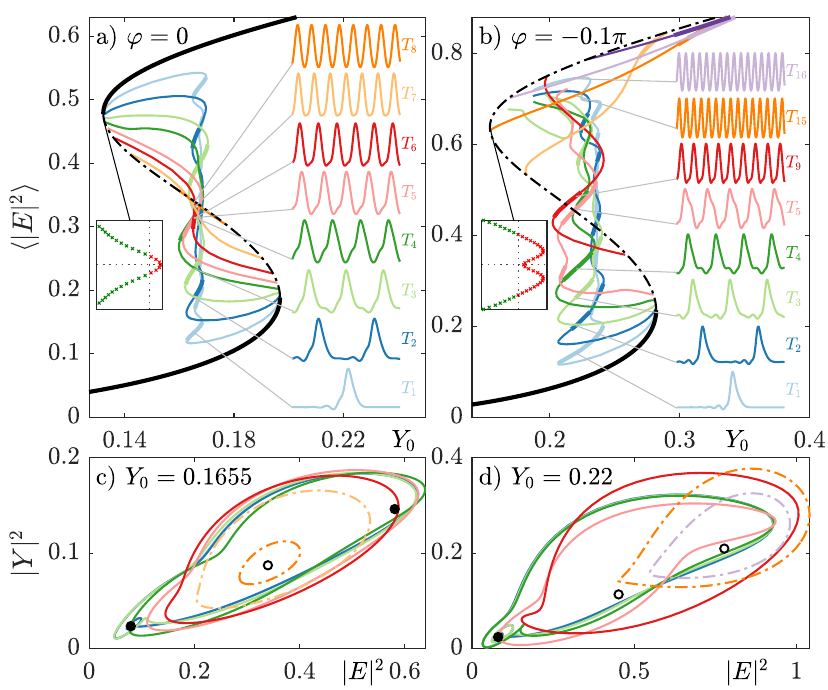}
	\caption{ (a) and (b) Bifurcation diagram of Eqs.~(\ref{eq:DAE1}) and (\ref{eq:DAE2}) obtained for $\tau=50$ for $\varphi=0$ and $\varphi=-0.1\pi$, respectively. The HSS and the
TLS branches are depicted in black and colored lines. On the right,
solution profiles along the branch indicated by the gray lines are plotted. The insets show the eigenvalue spectrum close to the left fold of the HSS branch. (c) and (d): Projection of the phase space onto the $(|E|^2,|Y|^2)$ plane in which the trajectories from the profiles in (a) and (b) are plotted in the same colors. Solid (dashed-dotted) and thick (thin) lines are stable (unstable) solutions, respectively. Other parameters as in Fig. \ref{fig:stability_plane}.}
	\label{fig:onset_turing}
\end{figure} 
Inside the region bounded by the fold lines, the system exhibits a bistable response. It was demonstrated in~\cite{SPV-OL-19,SJG-OL-22} that in the normal dispersion regime, bright and dark TLSs appear for $\varphi=0$ via the locking of domain walls connecting the high and low HSSs. Due to the oscillatory tails induced by the cavity dispersion~\cite{SCM-PRL-19}, these TLSs can interlock at multiple equilibrium distances leading to a rich ensemble of multistable solutions. Note that close to the onset of bistability located at $\varphi\approx0.15\pi$, $\tilde\varphi\approx0$ yielding the linear cavity response. There, the dynamics is governed by the LLE with third-order dispersion~\cite{SGJ-PRL-22}.

The TLS regime persists qualitatively up to a phase of $\varphi\approx-0.05\pi$ from where on the upper HSS becomes Turing unstable as indicated by the emerging yellow region in Fig.~\ref{fig:stability_plane}(a). The transition between the two regimes is illustrated in Fig.~\ref{fig:onset_turing}~(a,b) where bifurcation diagrams in $Y_0$ for $\varphi=0$ and $\varphi=-0.1\pi$, respectively, are presented. Here, not only the TLSs with a periodicity of $\tau$ are shown but also the higher harmonic solutions with shorter periods such that the TLS repeats multiple times per round trip. These solutions originate in the various AH bifurcations that occur when the pseudo-continuous spectrum crosses the imaginary axis. Due to the scaling of the eigenvalues on the pseudo-continuous spectrum of $2\pi/\tau$, the emerging periodic solutions possess a periodicity of $T_n\approx\frac{\tau}{n}, n\in\mathbb{N}$. Note that for the square waves, the spectrum does not have a leading zero eigenvalue resulting in the possible periods of $T_n^{SW}\approx\frac{2\tau}{2n-1}$.

For $\tau=50$ and $\varphi=0$, we find TLSs up to $T_8$  while for $\varphi=-0.1\pi$ we find solutions up to $T_{17}$ while only $T_{1-5},T_9, T_{13}$ and $T_{15-17}$ are depicted in Fig.~\ref{fig:onset_turing}(b) for clarity. The right insets show exemplary solution profiles of bright TLSs over the length of $\tau$ for $Y_0=0.1655$ ($\varphi=0$) and $Y_0=0.22$ ($\varphi=-0.1\pi$), respectively. The occurrence of higher frequencies for $\varphi=-0.1\pi$ can be explained by the change of shape of the pseudo-continuous spectrum which is depicted in the left insets in Figs.~\ref{fig:onset_turing}(a) and \ref{fig:onset_turing}(b), where it is shown close to the upper fold. Here one can see the transition from a uniform spectrum with a leading zero eigenvalue to a Turing unstable, i.e., a double well-like-shaped spectrum centered around a finite frequency. One can also observe a reordering of the AH bifurcation points: While for $\varphi=0$, the appearing branches are ordered as $T_1$ to $T_8$ and they disappear as $T_8$ to $T_1$ when following the HSS branch, for $\varphi=-0.1\pi$ this is not the case. The fact that in Fig.~\ref{fig:onset_turing}(b) the $T_1$ solution does not connect back to the HSS is because many more branches exist for the given parameters but are left out because they are unstable.

Figures \ref{fig:onset_turing}(c) and \ref{fig:onset_turing}(d) show a projection onto the $(|E|^2,|Y|^2)$-plane in which the trajectories of the exemplary bright TLSs in Figs.~\ref{fig:onset_turing}(a) and \ref{fig:onset_turing}(b) are plotted. Here, solid and dashed-dotted lines correspond to stable and unstable orbits, respectively. Further, the filled and open circles mark the position of the stable and unstable HSSs. Notably, the trajectories of the $T_{1-3}$ solutions are almost undistinguishable since the size of the domain is much bigger than the size of the individual structures. These trajectories consist in an excursion close to the upper HSS that connects to the lower HSS. These are periodic orbits that converge towards a \emph{ homoclinic} connection in the limit of infinite delay $\tau$ ~\cite{YRSW-PRL-19}. 
However, for higher harmonic solutions, the TLSs start to interact with each other which leads to a deformation of their temporal shapes as well as their phase space trajectories. These deformations can affect the stability: For $\varphi=0$, we find stable solutions up to $T_6$ while for $\varphi=-0.1\pi$ solutions remain stable up to $T_9$. The high-frequency solutions $T_{13-17}$ are all unstable in the bistable regime but can be stabilized for higher injection.
\begin{figure}[t!]
	\centering
	\includegraphics[width=1\columnwidth]{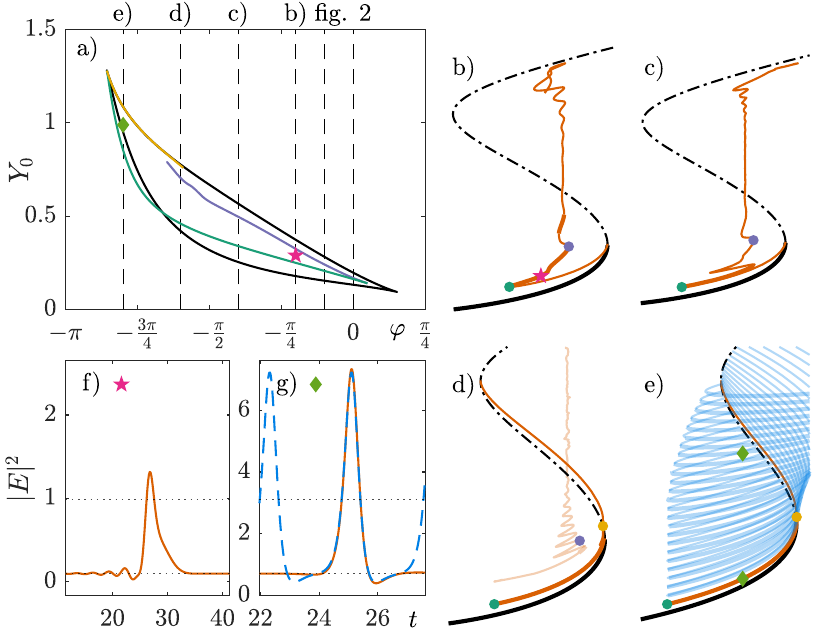}
	\caption{(a) Two-parameter bifurcation diagram in the $(\varphi, \,Y_0)$-plane. The black and colored lines correspond to the folds of the HSS and TLS solutions, respectively. (b)--(e) Cuts in the $Y_0$-direction for $\varphi=\{-0.2\pi,-0.4\pi,-0.6\pi,-0.8\pi\}$ (marked by dashed lines in (a)) with the HSS solution in black and the TLS branch in orange. For e) the harmonic branches are shown in blue up to $T_{25}$. Thick (thin) lines correspond to stable (unstable) solutions. The green, yellow, and purple dots mark the fold bifurcations from panel (a) in the same colors. At the positions marked with a pink star in (b) and light green diamond in (e), the profiles are plotted in panels (f) and (g), respectively. Here, the dotted lines stand for the lower and upper HSS values. Other parameters as in Fig. \ref{fig:stability_plane}. }
	\label{fig:two_parameter_diag}
\end{figure} 

To further explore the regime of Turing unstable upper HSS solutions, we perform a two-parameter continuation in the $(Y_0,\,\varphi)$ plane, cf. Fig.~\ref{fig:two_parameter_diag}(a). Here, black lines stand for the folds' position of the HSS branch, whereas green and purple lines mark the position of the folds of the leading bright TLS. Further, dashed vertical lines mark cuts at different $\varphi$ values. The first two cuts correspond to the bifurcation diagrams presented in Fig.~\ref{fig:onset_turing}, whereas other cuts are shown in Figs.~\ref{fig:two_parameter_diag}(b) and \ref{fig:two_parameter_diag}(e). Here, the corresponding folds of the leading bright TLS shown in (a) are marked by dots in the same color. In panel (e), we also show the harmonic solution branches up to $T_{25}$. 

First, we notice that due to the unstable upper HSS, the stability on the TLS branch changes as well; see Fig.~\ref{fig:two_parameter_diag}(b). In particular, dark TLSs become unstable as they reside for a majority of their trajectory close to the unstable upper HSS value and hence experience the same instability. Here, only the TLSs on the first two stages of the snaking structure remain stable. Next, one can see that the shape of the TLS branch changes drastically if $\varphi$ is changed further; cf. panels (c) and (d). This can be explained as the TLS branch reconnects transcritically to another branch with the same period which is shown as a transparent orange line in Fig.~\ref{fig:two_parameter_diag}(d). In the course of the transition, the purple marked TLS fold moves to this branch, and a new fold, marked in yellow, appears on the TLS branch; cf. panel (a).

We notice in Fig.~\ref{fig:two_parameter_diag}(e) that the left TLS fold marked by the green dot moves out of the region of bistability. There, the corresponding bright TLS approaches \emph{a homoclinic solution} in contrast to the double heteroclinic connection of two domain walls for the other TLSs in the bistable region. Additionally, the peak intensity of the resulting TLS is much higher here compared to the bistable case in panel (b). This is shown in panels (f) and (g) where profiles marked by a pink star and light green diamond in panels (a), (b), and (e) are presented. Here, the dotted horizontal lines correspond to the lower and upper HSS values, respectively. The dashed blue line in panel (g) is the $T_{18}$ solution which is also marked by a green diamond in panel (e). We observe that profiles around the pulse are indistinguishable, and thus, the fundamental TLS can be interpreted to be an element of this higher frequency solution. The branches of the harmonic states shown in Fig.~\ref{fig:two_parameter_diag}(e) undergo a similar transformation as the fundamental solution from locked domain walls (cf. Fig.~\ref{fig:onset_turing}) toward homoclinic orbits. The resulting diagram resembles that of cavity solitons in the anomalous dispersion regime of the LLE. However, there, the branches are organized in a homoclinic snaking diagram whereas here, the branches are independent. 
Note that also higher frequencies are present, however the shape of these densely packed pulses is altered.
Further decrease in $\varphi$ results in a cusp for the folds of both the HSS and TLS around the same value of $\varphi=-0.86\pi$, showing that the region of existence of TLSs is still restricted by the folds along the HSS branch.

\begin{figure}[t!]
	\centering
	\includegraphics[width=1\columnwidth]{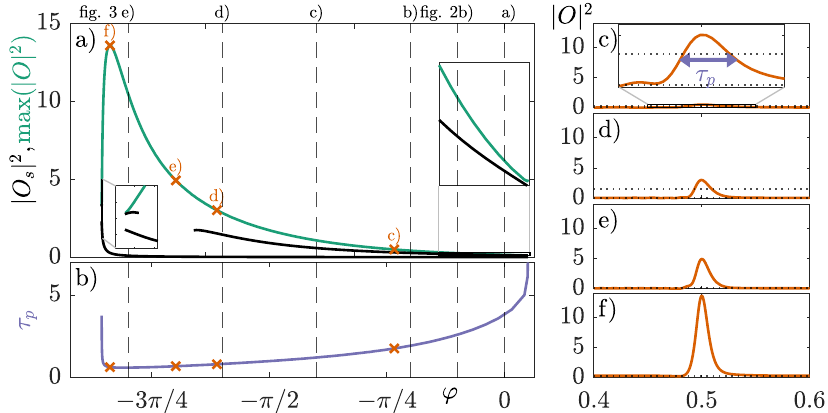}
	\caption{(a) Evolution of the shape of the TLS as a function of $\varphi$. The maximal intensity of the output field $O$ at the fold of the TLS branch (same data as in Fig. \ref{fig:two_parameter_diag}(a)) is shown in green, whereas black lines are the corresponding upper and lower HSS values at the TLS fold.
	The inset shows a zoom onto the region where the TLS branch emerges. (b) Pulse width $\tau_p$. Orange crosses mark position for which exemplary TLS profiles are shown in (c)-(f). Here, dotted lines are the lower and upper HSSs. The dashed lines in (a) and (b) mark the $Y_0$-cuts shown in Figs. \ref{fig:onset_turing}(a) and \ref{fig:onset_turing}(b) and \ref{fig:two_parameter_diag}~(b)--(e).}
	\label{fig:overshoot}
\end{figure}

To further analyze the transition of the bright TLS, we investigate the left fold of the TLS branch (green in Fig. \ref{fig:two_parameter_diag}) in more detail. Figure~\ref{fig:overshoot} shows this branch in green in the $(\varphi,\max(|O|^2))$-plane. Note that the HSS background on which this TLS resides changes with $\varphi$. Hence, we also plot the output intensity $|O_{s}|^2$ of the upper and lower HSS values corresponding to the injection value of the TLS-fold-branch (cf. black lines in Fig. \ref{fig:overshoot}(a)). Here, the point where the upper HSS line ends marks the value of $\varphi$ where the TLS branch leaves the bistable regime (cf. Fig. \ref{fig:two_parameter_diag}). Further, the dashed vertical lines are the cuts for Figs.~\ref{fig:onset_turing}(a) and \ref{fig:onset_turing}(b) and Figs.~\ref{fig:two_parameter_diag}(b) and \ref{fig:two_parameter_diag}(e).
In the right inset in Fig.~\ref{fig:overshoot}(a), one can see that close to $\varphi=0$, the maximal output intensity $|O|^2$ is close to the upper HSS value at the point of emergence since here the TLSs are composed of heteroclinic-like connections. Then, the maximal intensity grows steadily with decreasing $\varphi$ without encountering any discontinuity. It reaches a maximum around $\varphi\approx-0.83\pi$ before rapidly decreasing in intensity toward the lower HSS value. This happens around $\varphi\approx-0.86\pi$ where the TLS-fold-branch undergoes a cusp bifurcation. Note, that close to the cusp, the TLS-fold-branch reenters the bistable region (see left inset).
 As the pulse amplitude increases, its width $\tau_p$ defined as the full width at half maximum (FWHM) of $|O|^2-|O_{s}|^2$ decreases as depicted in Fig.~\ref{fig:overshoot}(b). Both characteristics of the profiles from panels (a) and (b) can be seen in Figs.~\ref{fig:overshoot}(c)--\ref{fig:overshoot}(f) where profiles at different points along the fold TLS branch are shown.

In conclusion, we have demonstrated that a KGTI, operated with normal dispersion and in the highly nonlinear regime where the mean-field approximation does not hold, exhibits a novel type of TLS that approaches a homoclinic orbit in the long delay limit. The latter are stable outside of the bistability region and are devoid of the shortcomings of double heteroclinic solitons induced by domain wall locking which should result in a more robust generation of OFCs.

\section*{Acknowledgment}
T.G.S. thanks the foundation “Studienstiftung des deutschen Volkes” for financial support,
J.J. and S.G. acknowledge the financial support of the projects KEFIR/AEI/10.13039/501100011033/ FEDER, UE, KOGIT, Agence Nationale de la Recherche (ANR-22-CE92-0009), Deutsche Forschungsgemeinschaft (DFG) via Grant Nr. 505936983 and Nr. 524947050.
\section*{Disclosures}
The authors declare no conflicts of interest


\begin{thebibliography}{10}
\newcommand{\enquote}[1]{``#1''}

\bibitem{CY-RMP-03}
S.~T. Cundiff and J.~Ye, \enquote{Colloquium: Femtosecond optical frequency
  combs,} {\protect\JournalTitle{Rev. Mod. Phys.}} \textbf{75}, 325--342
  (2003).

\bibitem{D-JOSAB-10}
S.~A. Diddams, \enquote{The evolving optical frequency comb,}
  {\protect\JournalTitle{J. Opt. Soc. Am. B}} \textbf{27}, B51--B62 (2010).

\bibitem{PPR-PR-18}
A.~Pasquazi, M.~Peccianti, L.~Razzari, \emph{et~al.}, \enquote{Micro-combs: A
  novel generation of optical sources,} {\protect\JournalTitle{Physics
  Reports}} \textbf{729}, 1 -- 81 (2018).

\bibitem{FB-CommPhys-19}
T.~Fortier and E.~Baumann, \enquote{20 years of developments in optical
  frequency comb technology and applications,}
  {\protect\JournalTitle{Communications Physics}} \textbf{2}, 153 (2019).

\bibitem{LCK-NAP-10}
F.~Leo, S.~Coen, P.~Kockaert, \emph{et~al.}, \enquote{Temporal cavity solitons
  in one-dimensional {K}err media as bits in an all-optical buffer,}
  {\protect\JournalTitle{Nat Photon}} \textbf{4}, 471--476 (2010).

\bibitem{HBJ-NAP-14}
T.~Herr, V.~Brasch, J.~D. Jost, \emph{et~al.}, \enquote{Temporal solitons in
  optical microresonators,} {\protect\JournalTitle{Nature Photonics}}
  \textbf{8}, 145--152 (2014).

\bibitem{WVM-LSA-17}
Z.~Wang, K.~Van~Gasse, V.~Moskalenko, \emph{et~al.}, \enquote{A iii-v-on-si
  ultra-dense comb laser,} {\protect\JournalTitle{Light: Science {\&}
  Applications}} \textbf{6}, e16260--e16260 (2017).

\bibitem{LL-PRL-87}
L.~A. Lugiato and R.~Lefever, \enquote{Spatial dissipative structures in
  passive optical systems,} {\protect\JournalTitle{Phys. Rev. Lett.}}
  \textbf{58}, 2209--2211 (1987).

\bibitem{CSY-PRL-10}
Y.~K. Chembo, D.~V. Strekalov, and N.~Yu, \enquote{Spectrum and dynamics of
  optical frequency combs generated with monolithic whispering gallery mode
  resonators,} {\protect\JournalTitle{Phys. Rev. Lett.}} \textbf{104}, 103902
  (2010).

\bibitem{CY-PRA-10}
Y.~K. Chembo and N.~Yu, \enquote{Modal expansion approach to
  optical-frequency-comb generation with monolithic whispering-gallery-mode
  resonators,} {\protect\JournalTitle{Phys. Rev. A}} \textbf{82}, 033801
  (2010).

\bibitem{CM-PRA-13}
Y.~K. Chembo and C.~R. Menyuk, \enquote{Spatiotemporal lugiato-lefever
  formalism for kerr-comb generation in whispering-gallery-mode resonators,}
  {\protect\JournalTitle{Phys. Rev. A}} \textbf{87}, 053852 (2013).

\bibitem{CRSE-OL-13}
S.~Coen, H.~G. Randle, T.~Sylvestre, and M.~Erkintalo, \enquote{Modeling of
  octave-spanning {K}err frequency combs using a generalized mean-field
  {L}ugiato--{L}efever model,} {\protect\JournalTitle{Opt. Lett.}} \textbf{38},
  37--39 (2013).

\bibitem{CPP-PRL-21}
L.~Columbo, M.~Piccardo, F.~Prati, \emph{et~al.}, \enquote{Unifying frequency
  combs in active and passive cavities: Temporal solitons in externally driven
  ring lasers,} {\protect\JournalTitle{Phys. Rev. Lett.}} \textbf{126}, 173903
  (2021).

\bibitem{LLK_OE_15}
V.~Lobanov, G.~Lihachev, T.~J. Kippenberg, and M.~Gorodetsky,
  \enquote{Frequency combs and platicons in optical microresonators with normal
  gvd,} {\protect\JournalTitle{Opt. Express}} \textbf{23}, 7713--7721 (2015).

\bibitem{XXL_NP_15}
X.~Xue, Y.~Xuan, Y.~Liu, \emph{et~al.}, \enquote{Mode-locked dark pulse {K}err
  combs in normal-dispersion microresonators,} {\protect\JournalTitle{Nature
  Photonics}} \textbf{9}, 594 EP -- (2015). Article.

\bibitem{PGL-OL-14}
P.~Parra-Rivas, D.~Gomila, F.~Leo, \emph{et~al.}, \enquote{Third-order
  chromatic dispersion stabilizes {K}err frequency combs,}
  {\protect\JournalTitle{Opt. Lett.}} \textbf{39}, 2971--2974 (2014).

\bibitem{PGK_OL_16}
P.~Parra-Rivas, D.~Gomila, E.~Knobloch, \emph{et~al.}, \enquote{Origin and
  stability of dark pulse {K}err combs in normal dispersion resonators,}
  {\protect\JournalTitle{Opt. Lett.}} \textbf{41}, 2402--2405 (2016).

\bibitem{PGG_PRA_17}
P.~Parra-Rivas, D.~Gomila, and L.~Gelens, \enquote{Coexistence of stable dark-
  and bright-soliton {K}err combs in normal-dispersion resonators,}
  {\protect\JournalTitle{Phys. Rev. A}} \textbf{95}, 053863 (2017).

\bibitem{CB-OL-17}
M.~Conforti and F.~Biancalana, \enquote{Multi-resonant lugiato\&\#x2013;lefever
  model,} {\protect\JournalTitle{Opt. Lett.}} \textbf{42}, 3666--3669 (2017).

\bibitem{I-OC-79}
K.~Ikeda, \enquote{Multiple-valued stationary state and its instability of the
  transmitted light by a ring cavity system,} {\protect\JournalTitle{Optics
  Communications}} \textbf{30}, 257 -- 261 (1979).

\bibitem{HW-JOSAB-15}
T.~Hansson and S.~Wabnitz, \enquote{Frequency comb generation beyond the
  {L}ugiato--{L}efever equation: multi-stability and super cavity solitons,}
  {\protect\JournalTitle{J. Opt. Soc. Am. B}} \textbf{32}, 1259--1266 (2015).

\bibitem{AWL-PRX-17}
M.~Anderson, Y.~Wang, F.~Leo, \emph{et~al.}, \enquote{Coexistence of multiple
  nonlinear states in a tristable passive {K}err resonator,}
  {\protect\JournalTitle{Phys. Rev. X}} \textbf{7}, 031031 (2017).

\bibitem{SPV-OL-19}
C.~Schelte, A.~Pimenov, A.~G. Vladimirov, \emph{et~al.}, \enquote{Tunable
  {K}err frequency combs and temporal localized states in time-delayed
  {G}ires-{T}ournois interferometers,} {\protect\JournalTitle{Opt. Lett.}}
  \textbf{44}, 4925--4928 (2019).

\bibitem{SJG-OL-22}
T.~G. Seidel, J.~Javaloyes, and S.~V. Gurevich, \enquote{A normal form for
  frequency combs and localized states in {K}err--{G}ires--{T}ournois
  interferometers,} {\protect\JournalTitle{Opt. Lett.}} \textbf{47}, 2979--2982
  (2022).

\bibitem{SGJ-PRL-22}
T.~G. Seidel, S.~V. Gurevich, and J.~Javaloyes, \enquote{Conservative solitons
  and reversibility in time delayed systems,} {\protect\JournalTitle{Phys. Rev.
  Lett.}} \textbf{128}, 083901 (2022).

\bibitem{DDEBT}
K.~Engelborghs, T.~Luzyanina, and D.~Roose, \enquote{Numerical bifurcation
  analysis of delay differential equations using dde-biftool,}
  {\protect\JournalTitle{ACM Trans. Math. Softw.}} \textbf{28}, 1--21 (2002).

\bibitem{KSG-OL-22}
E.~R. Koch, T.~G. Seidel, S.~V. Gurevich, and J.~Javaloyes,
  \enquote{Square-wave generation in vertical external-cavity
  kerr-gires-tournois interferometers,} {\protect\JournalTitle{Opt. Lett.}}
  \textbf{47}, 4343--4346 (2022).

\bibitem{GT-CRA-64}
F.~Gires and P.~Tournois, \enquote{Interferometre utilisable pour la
  compression d'impulsions lumineuses modulees en frequence,}
  {\protect\JournalTitle{C. R. Acad. Sci. Paris}} pp. 6112--6115 (1964).

\bibitem{SCM-PRL-19}
C.~Schelte, P.~Camelin, M.~Marconi, \emph{et~al.}, \enquote{Third order
  dispersion in time-delayed systems,} {\protect\JournalTitle{Phys. Rev.
  Lett.}} \textbf{123}, 043902 (2019).

\bibitem{VD-PRE-24}
A.~G. Vladimirov and D.~A. Dolinina, \enquote{Neutral delay differential
  equation model of an optically injected kerr cavity,}
  {\protect\JournalTitle{Phys. Rev. E}} \textbf{109}, 024206 (2024).

\bibitem{MB-JQE-05}
J.~Mulet and S.~Balle, \enquote{Mode locking dynamics in electrically-driven
  vertical-external-cavity surface-emitting lasers,}
  {\protect\JournalTitle{Quantum Electronics, IEEE Journal of}} \textbf{41},
  1148--1156 (2005).

\bibitem{CSV-OL-18}
P.~Camelin, C.~Schelte, A.~Verschelde, \emph{et~al.}, \enquote{Temporal
  localized structures in mode-locked vertical external-cavity surface-emitting
  lasers,} {\protect\JournalTitle{Opt. Lett.}} \textbf{43}, 5367--5370 (2018).

\bibitem{SHJ-PRAp-20}
C.~Schelte, D.~Hessel, J.~Javaloyes, and S.~V. Gurevich, \enquote{Dispersive
  instabilities in passively mode-locked integrated external-cavity
  surface-emitting lasers,} {\protect\JournalTitle{Phys. Rev. Applied}}
  \textbf{13}, 054050 (2020).

\bibitem{HGJ-OL-21}
D.~Hessel, S.~V. Gurevich, and J.~Javaloyes, \enquote{Wiggling instabilities of
  temporal localized states in passively mode-locked vertical external-cavity
  surface-emitting lasers,} {\protect\JournalTitle{Opt. Lett.}} \textbf{46},
  2557--2560 (2021).

\bibitem{Yanchuk2005}
S.~Yanchuk, \enquote{Properties of stationary states of delay equations with
  large delay and applications to laser dynamics,}
  {\protect\JournalTitle{Mathematical Methods in the Applied Sciences}}
  \textbf{28}, 363--377 (2005).

\bibitem{YRSW-PRL-19}
S.~Yanchuk, S.~Ruschel, J.~Sieber, and M.~Wolfrum, \enquote{Temporal
  dissipative solitons in time-delay feedback systems,}
  {\protect\JournalTitle{Phys. Rev. Lett.}} \textbf{123}, 053901 (2019).

\end{thebibliography}

\end{document}